\documentclass[prl,letterpaper,twocolumn,preprintnumbers,nofootinbib,superscriptaddress]{revtex4}

\usepackage{slashed}
\usepackage{amsmath,amssymb}
\usepackage{graphicx}
\usepackage{units}
\usepackage{bbold}
\usepackage{xcolor}
\usepackage{dsfont}
\usepackage[hyperfootnotes=false,colorlinks,citecolor=blue]{hyperref}

\usepackage{comment}
\usepackage[normalem]{ulem}     

\newcommand{\dd}{{\rm d}}
\newcommand{\ii}{{\rm i}}

\def \L {\mathcal{L}} 

\def \epsilon {\varepsilon} 

\def \vec#1{{\boldsymbol{#1}}}
\newcommand{\BR}{\ensuremath{\text{BR}}}
\newcommand{\hc}{\ensuremath{\text{h.c.}}}

\newcommand{\orcid}[1]{\href{https://orcid.org/#1}{#1}}

\begin{document}

\title{Nucleon Decays into Light New Particles in Neutrino Detectors}

\author{Julian Heeck}
\email[E-mail: ]{heeck@virginia.edu}
\thanks{\orcid{0000-0003-2653-5962}}
\affiliation{Department of Physics, University of Virginia,
Charlottesville, Virginia 22904, USA}

\author{Ian M. Shoemaker}
\email[E-mail: ]{shoemaker@vt.edu}
\thanks{\orcid{0000-0001-5434-3744}}
\affiliation{Center for Neutrino Physics, Department of Physics, Virginia Tech, Blacksburg, Virginia 24601, USA}

\begin{abstract}
Proton and neutron decays into light new particles $X$ can drastically change the experimental signatures and benefit from the complementarity of large water-Cherenkov neutrino detectors such as Super/Hyper-Kamiokande and tracking detectors such as JUNO and DUNE. The proton decays $p\to \ell^+ X$ and $p\to \pi^+ X$ with $m_X$ near phase-space closure lead to charged particles below Cherenkov threshold, rendering them practically invisible in Super- and Hyper-Kamiokande but not in JUNO and DUNE, which are therefore uniquely positioned for these baryon-number-violating signatures despite their smaller size. As an additional signature, such nucleon decays in Earth can produce a sizable flux of $X$ particles in underground detectors. We present a simple model in which nucleons decay into sub-GeV sterile neutrinos that subsequently decay through active-sterile neutrino mixing, with a promisingly large number of events in Super-Kamiokande even in the seesaw-motivated parameter space.
\end{abstract}

\maketitle

\section{Introduction}
\label{sec:intro}

Baryon number violation is  arguably the most sensitive probe we have of heavy physics beyond the Standard Model (SM)~\cite{FileviezPerez:2022ypk}. 
Several experiments are currently searching for such nucleon decays, with Super-Kamiokande (SK) dominating visible searches such as $p\to e^+\pi^0$~\cite{Super-Kamiokande:2020wjk} and low-threshold detectors such as SNO+ covering invisible final states such as $n\to 3\nu$~\cite{SNO:2022trz}.
New neutrino detectors are planned or already under construction that will aid these efforts: JUNO~\cite{JUNO:2015sjr}, Hyper-Kamiokande (HK)~\cite{Hyper-Kamiokande:2018ofw}, DUNE~\cite{DUNE:2020lwj}, and  THEIA~\cite{Theia:2019non}.

Light new particles $X$ can give rise to  non-standard nucleon decay modes such as $p\to \pi^+ X$ or $p\to e^+ X$, which could have fallen through the cracks in existing searches, and  have become an increasingly popular topic~\cite{Davoudiasl:2014gfa,Helo:2018bgb,McKeen:2020zni,Heeck:2020nbq,Fajfer:2020tqf,Heeck:2023soj,Fridell:2023tpb,Davoudiasl:2023peu,Domingo:2024qoj,Li:2024liy,Davoudiasl:2024xnq,Li:2025slp,Liao:2025vlj}.
Here, we study nucleon decays into new light neutral particles and emphasize the following points:
\begin{itemize}
    \item New particles with masses near the nucleon-decay threshold are accompanied by \textit{slow-moving} SM particles. If these are below Cherenkov threshold in water, SK/HK are effectively blind to such decays, allowing smaller detectors to set the best limits.
    \item Even nucleon-decay lifetimes in excess of $\unit[10^{33}]{yr}$  can yield a sizeable flux of $X$ particles sourced by the Earth. If the new particles $X$ are unstable, they could lead to displaced-vertex signatures~\cite{Domingo:2024qoj}. 
    
\end{itemize}
While we focus on two-body decays, our arguments also apply to more complex decays~\cite{Fridell:2023tpb, Li:2024liy, Liao:2025vlj}. 
We highlight a simple UV-complete example in which nucleons decay exclusively into light  sterile neutrinos, which then decay through their mixing with active neutrinos. A region of the seesaw-inspired parameter space is testable due to the novel nucleon-decay sterile-neutrino flux.

\section{Light bosons}

We will consider the simplest case~\cite{Heeck:2020nbq}: a new scalar $\phi$ with $B(\phi)=L(\phi)=1$, and focus on one particular $d=7$ interaction operator involving only right-handed fermions and  $\ell \in \{e,\mu\}$, using chiral perturbation theory~\cite{Claudson:1981gh,Nath:2006ut} 
to translate it into a hadronic Lagrangian:
\begin{align}
  \frac{\bar{u}^c d \bar{u}^c \ell \phi^*}{\Lambda^3_\ell} = -\frac{\beta}{\Lambda^3_\ell} \left(p- \frac{\ii p \pi^0}{\sqrt{2} f_\pi }- \frac{\ii n \pi^+}{ f_\pi }+\dots \right) \ell \phi^* \,,
  \label{eq:boson_operator}
\end{align}
where $\beta \simeq -\unit[0.013]{GeV^3}$ is a matrix element obtained via lattice QCD~\cite{Yoo:2021gql} and $f_\pi \simeq \unit[130]{MeV}$ is the pion decay constant that is used as a large expansion parameter to first order. The leading term is of the simple form $p \ell \phi^*$~\cite{McKeen:2020zni} and leads to the two-body proton decay~\cite{Heeck:2020nbq}
\begin{align}
\Gamma (p\to \ell^+ \phi) = \frac{|\vec{p}_\ell|}{16\pi}\frac{\beta^2}{\Lambda_\ell^6}\left(1+\frac{m_\ell^2}{m_p^2} - \frac{m_\phi^2}{m_p^2}\right) ,
\label{eq:p_to_phi}
\end{align}
with two-body final-state momentum $|\vec{p}_\ell| = |\vec{p}_\phi|$ in the proton rest frame, which is approximately the lab frame,
\begin{align}
        |\vec{p}_\ell|  = \frac{\sqrt{(m_p^2-(m_\ell+m_\phi)^2)(m_p^2-(m_\ell-m_\phi)^2) }}{2m_p}\,.
        \label{eq:momentum}
\end{align}
This decay mode is kinematically allowed for scalar masses $m_\phi < m_p - m_\ell$ for $\ell=e,\mu$.

UV completions of Eq.~\eqref{eq:boson_operator} 
either require $\phi$ to carry baryon number~\cite{Heeck:2020nbq,supplementary_material} or some fine-tuning to explain why the heavy particles behind~\eqref{eq:boson_operator} do not induce much faster $\phi$-less $d=6$ proton decays~\cite{Weinberg:1979sa}. In the former case,  $\phi$ is \textit{stable} -- making it an interesting dark-matter candidate but only leaving the mono-energetic anti-lepton to tag.
Giving up on $\phi$'s baryon number allows it to decay, e.g.~from mixing with the Higgs through the portal $\phi |H|^2$. One can even do away with the complex nature of $\phi$ and make it a (pseudo)Goldstone boson~\cite{Li:2024liy}, with typical decay channels into two photons or two leptons~\cite{Heeck:2017xmg}. For simplicity we will not consider these scenarios here.

\begin{figure}[tb]
    \centering
    \includegraphics[width=0.43\textwidth]{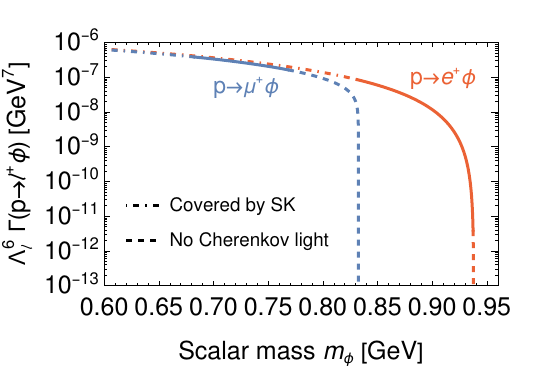}\vspace{-2ex}
    \caption{Free proton decay rates  vs $m_\phi$ from Eq.~\eqref{eq:p_to_phi}; the dot-dashed curve parts are covered by SK~\cite{Super-Kamiokande:2015pys};
    the dashed parts indicate  $\vec{p}_{\ell^+}$   below Cherenkov threshold in SK/HK.
    }
    \label{fig:proton_decays}
\end{figure}

Limits on the decay mode from Eq.~\eqref{eq:p_to_phi} have been obtained in SK~\cite{Super-Kamiokande:2015pys}  for $m_\phi = 0$ and exclude lifetimes $\Gamma^{-1}(p\to e^+ \phi) < \unit[8\times 10^{32}]{yr}$ and $\Gamma^{-1}(p\to \mu^+ \phi) < \unit[4\times 10^{32}]{yr}$, which probe effective scales $\Lambda_{e,\mu}\sim \unit[6\times 10^{9}]{GeV}$. 
The data and background events provided by SK~\cite{Super-Kamiokande:2015pys} for  $e^+$ ($\mu^+$) momenta above $\unit[100]{MeV}$ (\unit[200]{MeV}) can be used to obtain similar lifetime limits for $m_\phi \lesssim \unit[0.83]{GeV}$ ($\unit[0.68]{GeV}$). A dedicated search could probe even larger $m_\phi$, until the lepton momentum~\eqref{eq:momentum} is at last below the Cherenkov threshold of $1.14 m_\ell$~\cite{Heeck:2019kgr}, 
which occurs for
\begin{align}
\unit[937.5]{MeV} &\lesssim m_\phi \lesssim \unit[937.8]{MeV}\,, \quad \text{ for } \ p \to e^+ \phi\,,\\
\unit[768.2]{MeV} &\lesssim m_\phi \lesssim \unit[832.6]{MeV}\,, \quad \text{ for } \ p \to \mu^+ \phi\,,\end{align}
see also Fig.~\ref{fig:proton_decays}. The practical lower bound on $m_\phi$ to generate Cherenkov rings in SK could well be much lower.

SK is effectively blind to free-proton decays in these $m_\phi$ ranges, the only signatures being scattering of these slow $\ell^+$ in the detector, $e^+ e^-$ annihilation in the positron case,  as well as the eventual $\mu^+$ decay in the muon case, although these could happen outside of the detector. 
For the protons inside SK's oxygen nuclei, Fermi motion can effectively boost the $\ell^+$ and push it above the Cherenkov threshold, but SK's sensitivity will still be drastically reduced.
A more realistic limit in this case comes from invisible proton decay searches, which reach $\unit[10^{30}]{yr}$~\cite{SNO:2022trz} and can likely be improved by at least an order of magnitude in JUNO~\cite{JUNO:2024pur}, although the analysis has only been performed for invisible \textit{neutron} decay so far.
Dedicated searches for $p\to \ell^+ \phi$ in these slow-lepton regions in JUNO should then at least reach $\unit[10^{31}]{yr}$, likely more, given the additional mono-energetic lepton to tag; the larger volume and excellent track reconstruction of DUNE should push these searches even further, making $p\to \ell^+ \phi$ a rare example of a proton decay channel that is best searched for in JUNO \& DUNE, despite their smaller size compared to SK/HK.
We encourage sensitivity studies by our experimental colleagues to ascertain the actual reach, given that detector efficiencies and background play a major role here.

\section{Light fermions}

We  extend the SM by a new gauge-singlet Dirac fermion $\chi$ with  $B(\chi)=1$ and effective  couplings~\cite{delAguila:2008ir}
\begin{align}
&\frac{(\overline{u}_R^c d_R)(\overline{d}_R^c \chi_L^c)}{\Lambda_1^2}+\frac{(\overline{Q}_L^c Q_L)(\overline{d}_R^c \chi_L^c)}{\Lambda_2^2}  \label{eq:sterile_neutron_operators}\\
&\quad =\left(\frac{\beta}{\Lambda_1^2}+\frac{\alpha}{\Lambda_2^2}\right) \overline{\chi} P_R \left( n + \frac{\ii \pi^0 n}{\sqrt{2} f_\pi}  - \frac{\ii \pi^- p}{ f_\pi}+ \dots \right)    , \nonumber
\end{align}
with $\alpha \simeq -\beta$~\cite{Brodsky:1983st,Gavela:1988cp,Yoo:2021gql}. Neglecting heavier hadrons allows us to combine both operators into one effective parameter, $\epsilon\equiv \frac{\beta}{\Lambda_1^2}+\frac{\alpha}{\Lambda_2^2}$, which is a (small) mass mixing term between the right-handed neutron and the \textit{sterile neutron} $\chi$ that can also be traded for a mixing angle~\cite{Cline:2018ami}.

\begin{figure}[tb]
    \centering
    \includegraphics[width=0.43\textwidth]{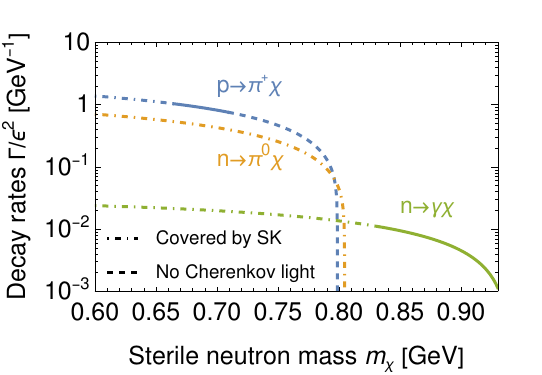}\vspace{-2ex}
    \caption{Free nucleon decay rates from Eq.~\eqref{eq:sterile_neutron_operators} vs $m_\chi$; the dot-dashed curve parts  are covered by SK~\cite{Super-Kamiokande:2013rwg,Super-Kamiokande:2015pys};
    the dashed part  indicates  $\vec{p}_{\pi^+}$  below Cherenkov threshold in SK/HK.
    }
    \label{fig:neutron_decays}
\end{figure}

The operator~\eqref{eq:sterile_neutron_operators} is to be supplemented with the usual hadron Lagrangian, including the neutron's magnetic moment term~\cite{Fornal:2018eol}, and can then be used to calculate nucleon decays after rotating $n$ and $\chi$ to the mass basis:
\begin{align}
\Gamma (n\to \chi \gamma) &= \frac{\epsilon^2 e^2 g_n^2 }{16\pi m_n} \left(1-\frac{m_\chi^4}{m_n^4}\right) ,\\
\Gamma (n\to \chi \pi^0) &\simeq \tfrac12 \Gamma (p\to \chi \pi^+) \simeq\frac{\epsilon^2 (1+g_A)^2 m_n}{64\pi f_\pi^2} \,,
\end{align}
with $g_n\simeq -3.83$,  $g_A \simeq  1.27$, and $m_{\pi,\chi}\to 0$ for the $\pi$ modes due to the lengthy expression, see Ref.~\cite{Davoudiasl:2014gfa} (Fig.~\ref{fig:neutron_decays}).

For small $m_\chi$, $p\to \chi \pi^+$ is the dominant nucleon decay rate, with $n\to \chi \pi^0$ an isospin-factor of 2 smaller~\cite{Helo:2018bgb}. For $m_\chi\to 0$, we can apply limits from SK's $N\to \pi \nu$ search~\cite{Super-Kamiokande:2013rwg}, which are $\unit[3.9\times 10^{32}]{yr}$ for protons and $\unit[1.1\times 10^{33}]{yr}$ for neutrons, the latter giving  \textit{slightly} better limits on $\epsilon$: $\epsilon < \unit[4\times 10^{-33}]{GeV}$, probing scales $\Lambda_{1,2} \sim \unit[2\times 10^{15}]{GeV}$. For $m_\chi > 0$, the $p\to \chi \pi^+$ signature becomes even more challenging as the reconstruction efficiency decreases~\cite{Super-Kamiokande:2013rwg}, until the $\pi^+$ eventually falls entirely below the Cherenkov threshold for $\unit[0.71]{GeV} \lesssim m_\chi < m_p-m_{\pi^+} $,  making it invisible in SK. The $\mu^+$ from $\pi^+$ decay is below threshold, too, leaving only the odd Michel positron as a signal in SK. In this mass range, JUNO and DUNE are poised to be the most sensitive detectors for $p\to  \pi^+\chi$, similar to $p\to \ell^+\phi$.

However,  this argument does not apply to the isospin-related mode $n\to \chi \pi^0$, as the outgoing $\pi^0\to \gamma\gamma$ photons always carry away at least $m_{\pi^0}/2$ in momentum and thus remain visible up until the phase-space closure $m_\chi \sim m_n - m_{\pi^0}$ (Fig.~\ref{fig:neutron_decays}). SK's reconstruction efficiency actually \textit{increases} for slow $\pi^0$~\cite{Super-Kamiokande:2013rwg}, so limits on $n\to \chi \pi^0$ are likely already of order $\unit[10^{33}]{yr}$ over the entire $\chi$ mass range.

For $ m_n - m_{\pi^0} < m_\chi < m_n$, the dominant process is $n\to \gamma\chi$~\cite{Davoudiasl:2014gfa} (Fig.~\ref{fig:neutron_decays}). 
The SK limit $\Gamma^{-1}(n\to\nu\gamma) < \unit[5.5\times 10^{32}]{yr}$~\cite{Super-Kamiokande:2015pys} likely approximately applies to $n\to \gamma\chi$ with $|\vec{p}_\gamma|> \unit[100]{MeV}$, or $m_\chi < \unit[0.83]{GeV}$.
For larger $m_\chi$, the limits eventually drop to the invisible-neutron case,  $\unit[10^{30}]{yr}$~\cite{SNO:2022trz}, and then by many orders of magnitude once bound-nucleon decays become kinematically forbidden for $m_\chi \gtrsim \unit[937.993]{MeV}$~\cite{Fornal:2018eol,McKeen:2020zni}.

In this light-fermion setup, JUNO 
\& DUNE should be able to outperform SK/HK in $p\to \chi \pi^+$ for $m_\chi\gtrsim\unit[0.7]{GeV}$, although most models~\cite{Helo:2018bgb} would be better constrained via the isospin-related Cherenkov-friendly $n\to \chi \pi^0$.\footnote{The two channels \textit{can} be decoupled in the parameter-space region $\frac{\beta}{\Lambda_1^2}+\frac{\alpha}{\Lambda_2^2}\simeq 0$, where isospin-breaking effects due to $m_d\neq m_u$ cannot be ignored.} Dedicated sensitivity studies are necessary to identify the best detector for $n\to \gamma\chi$ with $m_\chi \gtrsim \unit[0.83]{GeV}$, which blends into invisible-neutron territory.\vspace{-2ex}

\section{A simple model}\vspace{-2ex}

\begin{figure}[tb]
    \centering
    \includegraphics[width=0.45\textwidth]{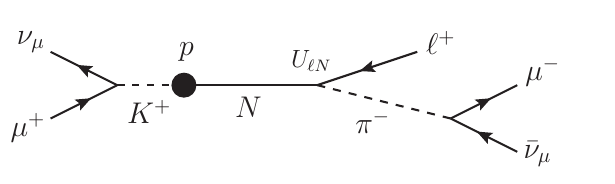}\vspace{-3ex}
    \caption{Proton decay chain $p\to K^+ N, N\to \pi^-\ell^+$. Majorana neutrinos $N$ can also decay to $\pi^+\ell^-$. Detectors might see parts of this chain depending on  the $N$ decay length.
    }
    \label{fig:displaced_proton_decay}
\end{figure}

As a simple realization of the above sterile-neutron setup~\cite{Fajfer:2020tqf}, we extend the SM by the scalar leptoquark $\bar{S}_1\sim (\bar{\vec{3}},\vec{1},-2/3)$ and several right-handed neutrinos $N$:
\begin{align}
\hspace{-1ex}    \lambda_{ ab} \overline{u}^c_{a} \bar{\mathcal{S}}_1 N_{ b} +
  \xi_{ ab} \overline{d}^c_{a} \bar{\mathcal{S}}_1^{* } d_{b} 
  +y_{ab} \overline{L}_a \tilde{H} N_{ b} +\frac{m_{ab} }{2} \overline{N}^c_a N_b\,,
  \label{eq:simple_N_model}
\end{align}
with generation indices $a, b$. 
Without the $N$, baryon number is conserved upon assigning $B(\bar{S}_1)=2/3$, so all nucleon decays \textit{must} involve $N$, without having to impose any symmetries as before. But without any additional quantum numbers, $N$ will decay through the Higgs Yukawa coupling $y$, giving rise to the decay chain of Fig.~\ref{fig:displaced_proton_decay}.
$N$ plays the dual role of sterile neutron \textit{and} sterile neutrino, with well-studied decay channels parametrized by their mixing $U_{\ell N}$ with SM neutrinos $\nu_\ell$~\cite{Atre:2009rg,Bondarenko:2018ptm,Coloma:2020lgy}.  
We  assume $N$ to be Majorana fermions here, so they give rise to seesaw neutrino masses~\cite{Minkowski:1977sc,Mohapatra:1979ia,Yanagida:1979as,Gell-Mann:1979vob}, generically of order $m_\nu \sim |U_{\ell N}|^2 m_N$. This also gives rise to $\Delta B = 2$~\cite{Proceedings:2020nzz} neutron--antineutron conversions, which are however sub-leading compared to nucleon decays in our region of interest. 
The physical picture barely changes for Dirac $N$, except that  Eq.~\eqref{eq:simple_N_model} then conserves $U(1)_{B-L}$~\cite{Heeck:2023soj} so the decay chain is restricted to $p\to \bar N \to $~anti-lepton (Fig.~\ref{fig:displaced_proton_decay}).

The down-quark couplings are antisymmetric in flavor space, $\xi_{ab} = - \xi_{ba}$, so the dominant nucleon decay for $m_N < m_p - m_K$ is $p\to K^+ N$. This signature was recently studied in Ref.~\cite{Domingo:2024qoj}, where $\bar{S}_1$ was identified with a supersymmetric quark partner. Since the kaon is already below Cherenkov threshold for $m_N = 0$, the SK limit $\Gamma^{-1}(p\to\nu K^+) < \unit[5.9\times 10^{33}]{yr}$~\cite{Super-Kamiokande:2014otb} essentially applies for all $m_N < m_p - m_K$~\cite{Domingo:2024qoj}.
For $m_p-m_K < m_{N}<m_p-m_\pi$, the proton can still decay to $ \pi^+ N$, albeit $G_F$ suppressed~\cite{Fajfer:2020tqf}, and the nucleon-decay discussion follows the sterile-neutron case from above.

In addition to the aforementioned nucleon decay channels, which can be complementarily covered by SK/HK, JUNO, and DUNE, the model from Eq.~\eqref{eq:simple_N_model} also provides $N$ decays. Nucleon decays such as  $p \to K^{+} N$ in Earth generate a flux of quasi-mono-energetic sterile neutrinos $N$ with decay length 
$
    \ell_{D} = \tau_N |\vec{p}_N|/m_N .
$
If $N$ can be produced and detected within the detector, it is  a displaced-vertex signature~\cite{Domingo:2024qoj} (Fig.~\ref{fig:displaced_proton_decay}),   with number of $N$ decays
\begin{equation}
N_{{\rm decays}}^\text{in} = T~\Gamma_{p\rightarrow N} N_p^\text{det}    \left( 1 -e^{- \Delta L  /\ell_{D}} \right),
\end{equation}
where $ N_p^\text{det} $ is the number of protons inside the detector, $T$ the measurement time, and $\Delta L$ the  effective detector length. Such displaced-vertex events, in which the entire decay chain of Fig.~\ref{fig:displaced_proton_decay} occurs inside the detector, require a rather small $\ell_D \lesssim \Delta L$. Existing limits on the mixing angles $U_{\ell N}$~\cite{Bolton:2019pcu} mostly forbid such fast $N$ decays, except in the $U_{\tau N}$ case. In the narrow region of parameter space in which $N$ has a short-enough decay length -- e.g.~because it is produced almost at rest -- the actual decay \textit{time} always exceeds $\unit[10^{-5}]{s}$ ($\unit[10^{-4}]{s}$ for $U_{eN}$), a considerable delay.

\begin{figure}[tb]
    \includegraphics[width=0.45\textwidth]{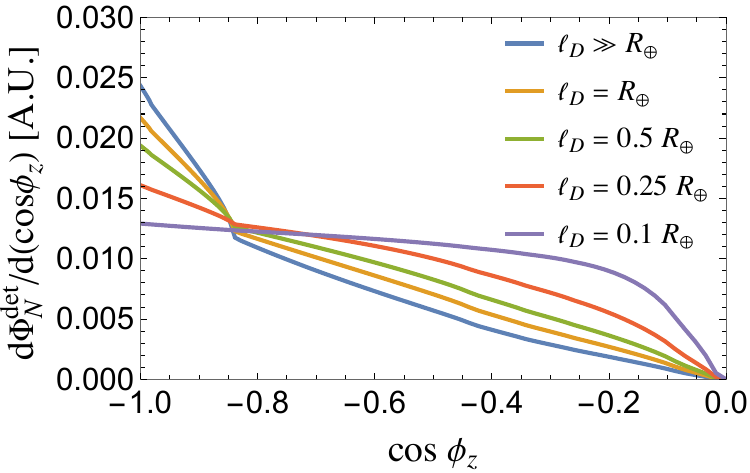} 

    \caption{ The zenith-angle $\phi_z$ distribution of the sterile neutrino flux at the detector for a few representative decay lengths $\ell_D$. For $\ell_D\ll R_\oplus$, the flux becomes increasingly isotropic and eventually extends to $0< \cos\phi_z$ for $\ell_D\,$overburden. The abrupt change in slope comes from the sharp density change near the Earth's core in the PREM~\cite{Dziewonski:1981xy}. 
    }
    \label{fig:angular}
\end{figure}

The more relevant number of $N$ decays inside a detector is produced by proton decays \textit{outside} the detector for $\mathcal{O}(\Delta L) \lesssim \ell_{D} \lesssim  \mathcal{O}(2 R_{\oplus})$. For a detector located near the Earth's surface, $\vec{R}_{\oplus}$, we can obtain the $N$ flux $ \Phi_{N}^\text{det}$ by adding up all the contributions from the protons in each small volume, $ n_{p}(r) \dd^{3}r$, for  number density $n_{p}$, over all possible proton decay locations $\vec{r}$  within the Earth,
\begin{equation}
    \Phi_{N}^\text{det} = \frac{\Gamma_{p\rightarrow N}}{4\pi} \int \dd^{3} r~\frac{n_{p}(r)}{|\vec{R}_{\oplus}- \vec{r}|^{2} }~e^{-|\vec{R}_{\oplus}-\vec{r}|/\ell_{D}}\,,
    \label{eq:outside_flux}
\end{equation}
where $|\vec{R}_{\oplus}-\vec{r}| = \sqrt{R_{\oplus}^{2} + r^{2} - 2rR_{\oplus} \cos \theta}$, and $\theta$ the angle between the vectors $\vec{r}$ and $\vec{R}_{\oplus}$.
For a uniform proton density $n_p = N^p_\oplus/(4\pi R_\oplus^3/3)$, we can solve the integral to
\begin{align}
    \Phi_{N}^\text{det} = \frac{\Gamma_{p\rightarrow N} }{2} n_p\,
     \ell_D\left[1-\frac{\ell_D}{2 R_\oplus} \left(1- e^{-2 R_\oplus/\ell_D}\right)\right] .
\end{align}
The number of $N$ decays within the detector volume is
\begin{equation}
    N_\text{decays}^\text{out} = \Phi_{N}^\text{det}(A_{{\rm eff}} T) \left(1 - e^{-\Delta L/\ell_{D}}\right),
    \label{eq:Nout}
\end{equation}
where $A_\text{eff}$ is the effective area of the detector.
This is maximal in the decay-length region $\Delta L \lesssim \ell_D \lesssim R_\oplus$, with 
\begin{align}
     \text{max}( N_\text{decays}^\text{out} ) &=\tfrac12 \Gamma_{p\rightarrow N} n_p A_\text{eff} \Delta L\, T \\
      &\simeq 5 \left(\frac{\unit[10^{33}]{yr}}{\tau_{p\to N}} \right)\left(\frac{ A_\text{eff} \Delta L\, T}{(\unit[10]{m})^3\unit[10]{yr}} \right) .
      \label{eq:Nmax}
\end{align}
The factor 1/2 is due to the assumption that the detector is at the surface, receiving only flux from below. Once $\ell_D$ is smaller than the overburden, $\mathcal{O}(\unit[1]{km})$ for SK, the flux is  doubled; once $\ell_D <\Delta L$, $N_\text{decays}^\text{in}$ should be added.
The direction of the $N$ flux is determined by the decay length: for $\ell_D \lesssim \unit{km}$, $N$ arrives isotropically, although this region of parameter space requires large $U_{\ell N}$ and is strongly constrained. For decay lengths $\sim R_\oplus$, $N$ are mostly coming from the opposite side of Earth, see Fig.~\ref{fig:angular} for the zenith-angle $\phi_z$ distribution~\cite{Fields:2004tf}, related to $\theta$ via
\begin{equation}
    \frac{\sin \phi_{z}}{r} = \frac{\sin\theta}{\sqrt{R^{2} +r^{2} - 2rR \cos \theta}} \,.
\end{equation}

For SK (HK), we use $\Delta L \simeq \unit[32]{m}$ ($\unit[67]{m}$) and $A_\text{eff} \simeq \unit[707]{m^2}$ ($\unit[3421]{m^2}$) in our analysis, not taking SK's fiducial volume extension~\cite{Super-Kamiokande:2020wjk} into account, and $T= \unit[20]{yr}$ for both. DUNE is in-between SK and HK.  
SK already exceeds the benchmark numbers~\eqref{eq:Nmax} for the space-time volume $A_\text{eff} \Delta L\, T$  and could have up to 5 $N$ decays in their detector even for a proton decay lifetime of $\unit[10^{35}]{yr}$! 

\begin{figure*}
    \includegraphics[width=0.32\textwidth]{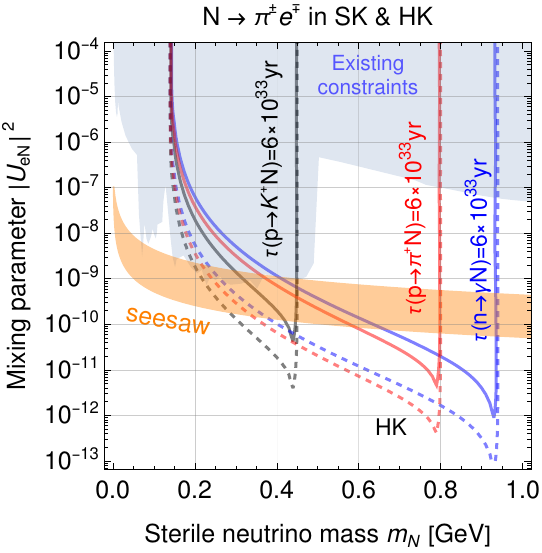} 
    \includegraphics[width=0.32\textwidth]{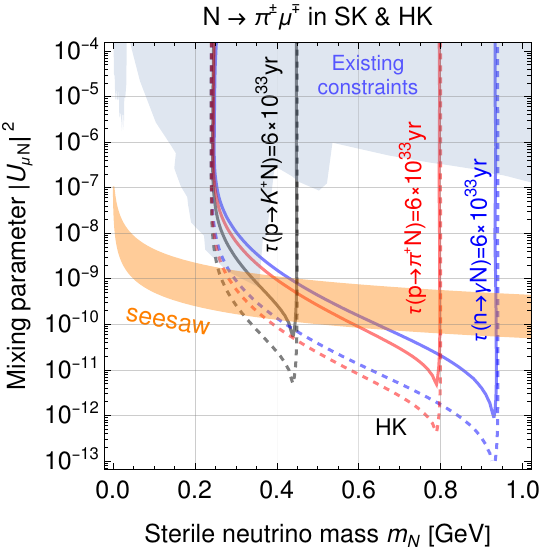}
    \includegraphics[width=0.32\textwidth]{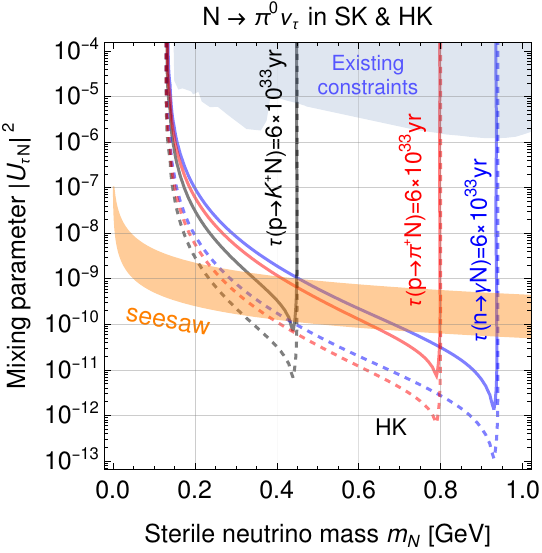}
    \caption{ Left: contours of $N_\text{sig}=5$ sterile neutrino decays $N\to\pi^\pm e^\mp$ inside SK (solid) and HK (dashed), produced by various proton decay channels with lifetime $\tau_{p\to N} =\unit[6\times 10^{33}]{yr}$. Also shown are existing lab constraints~\cite{Bolton:2019pcu} and the seesaw-motivated parameter space.
    Middle: same for $U_{\mu N}$ mixing and decay $ N\to \pi^\pm\mu^\mp$. 
    Right: for  $U_{\tau N}$ mixing and decay $ N\to \pi^0\nu_\tau,\pi^0\bar \nu_\tau$.
    }
    \label{fig:SuperK}
\end{figure*}

The number of $N$ decays into a given final state $X$ is then $N_\text{sig} =\BR(N\to X) ( N_\text{decays}^\text{in}+ N_\text{decays}^\text{out})$, with relevant branching ratios in our mass range given in Ref.~\cite{Coloma:2020lgy}.
We show contours of $N_\text{sig}=5$ for three different mixing scenarios in Fig.~\ref{fig:SuperK}, using the preliminary reference Earth model (PREM)~\cite{Dziewonski:1981xy} for the proton density, together with existing laboratory constraints~\cite{Bolton:2019pcu} and the generic seesaw expectation $m_\nu \sim |U_{\ell N}|^2 m_N$, with $m_\nu \in \{\unit[0.05]{eV},\unit[0.45]{eV}\}$, the lower bound coming from the atmospheric mass splitting, $\sqrt{\Delta m_\text{atm}^2}$, and the upper bound from KATRIN~\cite{KATRIN:2024cdt}.
For $U_{eN,\mu N}$, we focused on the $N\to\pi^\pm \ell^\mp$ decay, which has a large branching ratio and clean signature. For $U_{\tau N}$, no fully visible $N$ decay exists in the sub-GeV mass range, so we use $N\to \pi^0 \nu_\tau$. In all three cases going to smaller $m_N$ is possible but leaves us with  $N\to \ell^\pm \ell^\mp \nu$, a three-body decay involving missing energy.
The chosen production rate\footnote{We include the isospin-related neutron decays, $\Gamma (n\to K^0 N) = \Gamma (p\to K^+ N)$~\cite{Fajfer:2020tqf} and $\Gamma (n\to \pi^0 N) = \Gamma (p\to \pi^+ N)/2$~\cite{Helo:2018bgb}.}  $\tau_{p\to N}=\unit[6\times 10^{33}]{yr}$ corresponds to the estimated current limit for $p\to K^+ N$, but is overly pessimistic for $p\to \pi^+ N$ and especially $n\to \gamma N$, see discussion above, and could be at least a factor 5 smaller even in the SK covered mass regions.

The signature is a nearly mono-energetic flux of $N$; for very slow $N$, i.e.~near proton-decay phase-space closure, $N$ decays into back-to-back SM particles, while a more boosted $N$ emits the final states with a smaller opening angle and box-shape energy spectrum~\cite{Ibarra:2016fco,Garcia-Cely:2016pse}. 
For guidance: the blue SK contours in Fig.~\ref{fig:SuperK} roughly correspond to $\ell_D\sim \unit[3\times 10^4]{km}$, and scale with $|U_{\ell N}|^{-2}$.
These searches will face the atmospheric neutrino background, but given the nigh unconstrained proton-decay rates in some areas of parameter space the event numbers could be in the thousands. We encourage a dedicated search in SK, not least because it is a rare chance to probe the seesaw-motivated parameter space.

An additional $N$ flux is sourced by the Sun, but the  probability to decay in the detector is heavily suppressed. However,  satellites could see decays such as $N\to e^+e^-\nu$  in interplanetary space, analogous to Refs.~\cite{1981Natur.289..777T,Gustafson:2023hvm,Drewes:2024dem} but with heavier $N$. We leave this study for future work.

\section{Conclusions}
\label{sec:conclusions}

Baryon number violation, especially in the form of nucleon decays, has been long identified as a sensitive probe of physics beyond the SM. New light particles emitted in those decays can drastically change the experimental signatures and require dedicated analyses. Nucleon decays such as $p\to \ell^+ X$ or $p\to \pi^+ X$ could be hidden from the large water-Cherenkov detectors SK \& HK if $m_X$ is near phase-space closure, making them one of the few nucleon decays best searched for in the smaller detectors JUNO \& DUNE, illustrating their complementarity.
Nucleon decays in Earth also generate a potentially testable flux of $X$ particles that can again be studied in neutrino detectors, e.g.~through the sterile-neutrino decay chain $p\to K^+ N$, $N\to \pi^\pm \ell^\mp$, even for seesaw-suppressed active-sterile mixing angles. Our initial exploration of these signatures hints at many novel exciting opportunities for theoretical and experimental work that could lead to the groundbreaking discovery of new physics.

\section*{Acknowledgments}

We thank Diana Sokhashvili for discussions. 
This work was supported by a 4-VA at UVA Collaborative Research Grant and by  the U.S.~Department of Energy under Grant No.~DE-SC0007974 (JH) and DE-SC0020262 (IMS).

\bibliographystyle{utcaps_mod}
\bibliography{BIB.bib}

\appendix*

\section{Supplemental Material}

In this Supplemental Material, we collect example UV completions for the operators discussed in the letter for the interested reader.

Following Ref.~\cite{Heeck:2020nbq}, we provide an explicit UV completion for the operator $\bar{u}^c d \bar{u}^c \ell \phi^*/\Lambda^3_\ell$. We introduce the two scalars $\mathcal{S}_1$ and $\mathcal{S}_1'$, both in the  $SU(3)_C\times SU(2)_L\times U(1)_Y$ gauge representation $(\overline{\vec{3}},\vec{1},1/3)$, but carrying baryon number $-1/3$ and $2/3$, respectively, making $\mathcal{S}_1$ a leptoquark and $\mathcal{S}_1'$ a diquark, with relevant couplings
\begin{align}
    \L &= y_{u\ell}\mathcal{S}_1 \bar{u}^c \ell+y_{QL}\mathcal{S}_1 \bar{Q}^c L + y_{ud}\overline{\mathcal{S}}_1' \bar{u}^c d+ y_{QQ}\overline{\mathcal{S}}_1' \bar{Q}^c Q \nonumber\\
    &\quad+ \mu\,  \overline{\mathcal{S}}_1 \mathcal{S}_1' \phi^* +\hc ,
\end{align}
where $\mu$ is a coupling with dimension of mass. $\phi$ carries $B=1$ and lepton number can also be consistently assinged to all new particles. Integrating out the two heavy scalars gives rise to the $\Delta B$ operators
\begin{align}
   & \frac{\mu y_{ud}y_{u\ell}}{m_{\mathcal{S}_1}^2m_{\mathcal{S}_1'}^2}\bar{u}^c d \bar{u}^c \ell \phi^* +
    \frac{\mu y_{QQ}y_{u\ell}}{m_{\mathcal{S}_1}^2m_{\mathcal{S}_1'}^2}\bar{Q}^c Q \bar{u}^c \ell \phi^* \\
    &+ 
    \frac{\mu y_{ud}y_{QL}}{m_{\mathcal{S}_1}^2m_{\mathcal{S}_1'}^2}\bar{u}^c d \bar{Q}^c L \phi^* +
    \frac{\mu y_{QQ}y_{QL}}{m_{\mathcal{S}_1}^2m_{\mathcal{S}_1'}^2}\bar{Q}^c Q \bar{Q}^c L \phi^* \,.
\end{align}
This includes the desired operator $\bar{u}^c d \bar{u}^c \ell \phi^*/\Lambda^3_\ell$ with $\Lambda_\ell =\sqrt[3]{m_{\mathcal{S}_1}^2m_{\mathcal{S}_1'}^2/(\mu y_{ud}y_{u\ell})} $ as well as similar operators with different fermion chiralities, all of which lead to the same $p\to \ell^+\phi$ discussed in the main text. For $y_{QL}\neq 0$, this model also induces $n\to \bar{\nu}_\ell \phi$, subject to standard invisible-neutron constraints independent of $m_\phi$. If all Yukawa couplings are of order one and $m_{\mathcal{S}_1}\sim m_{\mathcal{S}_1'}\sim \mu$, nucleon decays probe scalar masses above $\unit[10^9]{GeV}$, high enough that one needn't worry about any other constraints, be it from colliders or precision measurements. Nucleon decays are uniquely positioned to probe this model as well as other models that lead to such $\Delta B $ operators. Of course, one can also consider pushing one of the scalars to TeV mass scales and reducing the Yukawa couplings appropriately to keep $\Lambda_\ell$ the same, in which case these particles \textit{could} show up at colliders or in other observables. Overall, there is vast parameter space for testable $p\to \ell^+\phi$.

UV completions for the other operators of interest in  the main letter, involving a light sterile fermion $\chi$,
\begin{align}
    \frac{(\overline{u}_R^c d_R)(\overline{d}_R^c \chi_L^c)}{\Lambda_1^2}+\frac{(\overline{Q}_L^c Q_L)(\overline{d}_R^c \chi_L^c)}{\Lambda_2^2}
\end{align}
are even simpler to construct. One UV completion has been provided in the main text, another again involves a scalar $\mathcal{S}_1\sim (\overline{\vec{3}},\vec{1},1/3)$, with generic couplings
\begin{align}
    \L &= y_{u\ell}\mathcal{S}_1 \bar{u}^c \ell+y_{QL}\mathcal{S}_1 \bar{Q}^c L+ y_{d\chi}\mathcal{S}_1 \bar{d}^c \chi^c\nonumber\\
     &\quad+ y_{ud}\overline{\mathcal{S}}_1 \bar{u}^c d+ y_{QQ}\overline{\mathcal{S}}_1 \bar{Q}^c Q +\hc
\end{align}
Assigning $\chi$ and $\mathcal{S}_1$ baryon numbers $+1$ and $2/3$, respectively, enforces $y_{u\ell} = y_{QL} = 0$ and thus only induces
\begin{align}
    \frac{y_{d\chi} y_{ud}}{m_{\mathcal{S}_1}^2}(\overline{u}_R^c d_R)(\overline{d}_R^c \chi_L^c)+\frac{y_{d\chi} y_{QQ}}{m_{\mathcal{S}_1}^2}(\overline{Q}_L^c Q_L)(\overline{d}_R^c \chi_L^c)
\end{align}
upon integrating out the heavy $\mathcal{S}_1$~\cite{Fornal:2018eol}. Without this choice, i.e.~for $y_{u\ell} \neq 0\neq y_{QL} $, one expects standard nucleon decays such as $p\to \pi^0 e^+$ and $n\to \pi^0 \bar{\nu}$ in addition to $p\to \pi^+ \chi$, but with unknown branching ratios.  
Once again there is plenty of model and parameter space that is best probed by nucleon decays, even more-so than in the previous example since we are now dealing with a dimension-\textit{six} operator, so the probed mass scale is even higher, easily of order $\unit[10^{15}]{GeV}$.

\end{document}